\documentstyle[multicol,aps,psfig]{revtex}

\def\beginwide{
        \end{multicols} \vspace*{-0.5cm} \noindent
        \rule{3.5in}{.1mm}\rule{.1mm}{5mm} \widetext \medskip }
\def\beginwidetop{
        \end{multicols} \vspace*{-0.5cm} \noindent
        \widetext \medskip }
\def\endwide{
        \hspace*{3.35in}~\rule[-5mm]{.1mm}{5mm}\rule{3.5in}{.1mm}
        \begin{multicols}{2} \vspace*{-1.0cm} \noindent }
\def\endwidebottom{
        \begin{multicols}{2} \vspace*{-1.0cm} \noindent }
\def\beq{\begin{equation}}
\def\eeq{\end{equation}}

\begin{document}

\title{\Large Depinning of a dislocation: the influence of 
long-range interactions}

\author{Stefano Zapperi$^1$ and Michael Zaiser$^{2}$}

\address{$^1$INFM, Universit\`a "La Sapienza",\\ P.le A. Moro 2
        00185 Roma, Italy\\
$^2$MPI f\"ur Metallforschung,\\
Heisenbergstr.1, D-70569 Stuttgart, Germany
}
\date{\today}
\maketitle

\begin{abstract}
\noindent The theory of the depinning transition of elastic manifolds in random media 
provides a framework for the statistical dynamics of dislocation systems 
at yield. We consider the case of a single flexible
dislocation gliding through a random stress field generated by a 
distribution of immobile dislocations threading through its 
glide plane. The immobile dislocations are arranged in a "restrictedly
random" manner and provide an effective stress field whose
statistical properties can be calculated explicitly. We write an equation
of motion for the dislocation and compute the associated depinning
force, which may be identified with the yield stress. Numerical simulations
of a discretized version of the equation confirm these results and 
allow us to investigate the critical dynamics of the pinning-depinning 
transition.
\end{abstract}


\section{Introduction}

The depinning transition of a dislocation moving in a glide plane has 
been investigated in the past in several papers using a line tension
approximation for the dislocation self-interaction and considering 
interactions with point obstacles \cite{SEV-91,DAN-97}. In fact, however, 
the bending of a dislocation line produces a long-range self-stress 
\cite{FOR-67,WIT-59} and the obstacles responsible for the pinning 
are often forest dislocations rather than point defects (i.e. solute atoms, 
small clusters). This distinction is important since dislocations, in addition 
to short-range interactions, give rise to long-range stresses in the glide 
plane and thereby cause a long-range correlated pinning field. Also the
'short-range' interactions (junction formation) with forest dislocations
which in real crystals contribute substantially to the pinning should not be 
modeled in terms of point obstacles, since the corresponding pinning forces 
exhibit spatial correlations with ranges of the order of the junction length, 
which is in turn of the same order as the obstacle (forest
dislocation) spacing \cite{SCH-72}. 

Since long-range interactions are known to affect the general scaling 
properties of the depinning transition \cite{KAR-98}, in this paper we 
study dislocation depinning considering explicitly the effects of nonlocal elasticity
and spatial correlations in the pinning field. To this end, we introduce a lattice model 
with long-range dislocation self-interactions that correctly recovers previous 
results for the bending of a dislocation segment. Using this model we show that 
the geometrical properties of the dislocation at depinning differ from those 
expected in line tension models, as shown by a different roughness exponent.

\section{Formulation of the model}

We consider a dislocation gliding in the $xy$ plane, parametrized by $\vec{R}(s)=(X(s),Y(s),0)$,
with Burgers vector $\vec{b}=(b,0,0)$, and interacting with a pinning field created by a random 
array of immobile dislocations threading through the plane. The resolved shear stress
acting locally on the dislocation is given by the sum of three contributions
\beq
\sigma_{xz}(\vec{R}) 
= \sigma_{xz}^{\rm ext}+\sigma_{xz}^{\rm s}(\vec{R})
+ \sigma_{xz}^{\rm p}(\vec{R}),
\eeq
where $\sigma_{xz}^{\rm ext}$ is the externally 
applied stress  and $\sigma_{xz}^{\rm p}$ is the pinning field.
We assume that electron and phonon drag lead to an overdamped 
motion of the dislocation, so that the velocity normal 
to the dislocation is given by $v_n= b\sigma_{xz}/\Gamma$, where $\Gamma$ is
an effective viscosity coefficient.
Using relations given by Hirth and Lothe \cite{HIR-68} for the stress 
fields of dislocation segments, the self-stress of the dislocation can 
be written as
\beq
\sigma_{xz}^{s}(\vec{R})=
\frac{b\mu}{4\pi} 
\int ds \frac{(Y-Y(s))\partial_s X
+(X-X(s))\partial_s Y/(1-\nu)}{|R-R(s)|^3}.
\eeq
where $\mu$ is the shear modulus and $\nu$ is Poisson's number.

To study the influence of long-range correlations in the pinning 
stress, we consider a stress field created by an array 
of immobile dislocations threading through the glide plane. We note
that it is not feasible to consider a fully random arrangment of 
the threading dislocations, as this would imply a diverging energy 
density of the pinning field \cite{WIL-69}. 
Hence some correlations in the dislocation arrangement creating
the pinning stress must be assumed to ensure a finite energy density. 
To this end we use Wilkens' construction of a 
'restrictedly random' distribution of dislocations \cite{WIL-69}
where the system is divided in boxes of side $l$, each of them containing $n$ dislocations with 
a fixed distribution of Burgers vectors placed at random within the box. For this model,
it is possible to compute analytically the statistical properties 
of the pinning field. In particular, the stress distribution 
is symmetric with tails decaying as $\rho/\sigma^3$, where $\rho = n/l^2$ 
is the density of dislocations creating the pinning field\cite{GRO-98}. 
The spatial correlation function of the pinning field can also be computed 
\cite{ZAI-00}. This function is given by
\beq
\langle\sigma_{xz}^{\rm p}(\vec{R}))\sigma_{xz}^{\rm p}(\vec{R}\;')\rangle=
C(\mu b)^2\rho f([\vec{R}-\vec{R'}]/\xi_{\rm s})\;,
\eeq
where the constant $C$ depends on the distribution of Burgers vectors and 
$\xi_{\rm s} = M/\sqrt{\rho}$ with $M \approx 0.25 \sqrt{n}$ is the 
correlation length of the pinning field. The 
correlation function $f(\vec{R}/\xi_{\rm s})$ has the following properties:
$f \to  - \ln(R/\xi_{\rm s})$ for $R = |\vec{R}| \to 0$ and
$f = 0$ for $R/\xi_{\rm s} > 4$.
The logarithmic singularity of $f$ at the origin is truncated
at the dislocation core radius $r_{\rm c} \approx b$, yielding a mean
square stress $\langle (\sigma_{xz}^{\rm p})^2 \rangle = 
\mu^2 b^2 C\rho \ln[\xi_{\rm s}/b]$. 
The spatial average of the pinning stress is, of course, 
zero. These results have been verified numerically.

\section{Depinning transition}

We consider a dislocation with average screw orientation (i.e., the average line
direction is parallel to the $x$ axis). In a first step, we perform a simplified 
analysis and assume that the dislocation has no overhangs (i.e $\vec{R}=(x,h(x),0)$),
and only small deformations (i.e. $|h-h'| \ll |x-x'|$). In this case, the equation 
of motion becomes
\beq{\Gamma\over b}\frac{\partial h}{\partial t}=\sigma_{xz}^{\rm ext}+
\frac{b\mu}{4\pi} \int\frac{dx}{|x-x'|^3}\left((h(x)-h(x'))+
\frac{(x-x')}{(1-\nu)}\frac{dh(x')}{dx'}\right)+\sigma_{xz}^{\rm p}(x,h). 
\label{eq:h}
\eeq
Eq. \ref{eq:h} describes an 'elastic' interface moving in
a quenched disordered medium. For this class of systems
we expect a depinning transition as a function of the 
external stress. The average velocity of the dislocation vanishes
in the long time limit when the external stress 
$\sigma\equiv\sigma_{xz}^{\rm ext}$
is below a critical value $\sigma_{\rm c}$, and scales as $(\sigma-\sigma_{\rm c})^\beta$ 
for $\sigma > \sigma_c$. The critical point may be identified with the yield 
stress.

We can obtain the dependence of the critical stress
on the density of the dislocations creating the pinning field by 
rescaling Eq. (\ref{eq:h}). If we define 
$\tilde x = [4\pi \sqrt{C\rho}] x$, $\tilde t = [4\pi C \rho
\mu b^2/\Gamma] t$ and $\tilde\sigma= [\sigma/(\mu b\sqrt{C\rho}]$,
the resulting equation depends only on $\nu$, $[M \sqrt{C}]$ and the rescaled
external stress, so that
\beq
\sigma_c = b\mu \sqrt{C\rho}~\tilde{\sigma}_{\rm c}(\nu,[M\sqrt{C}]),
\eeq
where $\tilde{\sigma}_{\rm c}$ is the depinning stress for the rescaled equation. 
When pinning is due to the cumulative effect of a large number of obstacles (weak 
pinning), this stress can be estimated using an argument which 
is originally due to Larkin \cite{DAN-97}.  On small scales and for 
weak pinning\footnote{In the context of dislocations
weak pinning corresponds to the case of Mott-Labusch statistics
\cite{KOC-75}.}, the self-interaction 
keeps the dislocation essentially straight. Hence the average force experienced
by a segment of length $L$ is $\sqrt{\langle (\sigma^{\rm p})^2 \rangle L \xi_{\rm s}}$.
Above a critical length $L_{\rm c}$ the dislocation can bow and adjust to the 
'contours' of the pinning field. Bowing a near-screw segment of length $L$ over a 
characteristic distance $\xi _{\rm s} \ll L$ requires a characteristic
force $\sim \mu b (1+\nu)\xi_{\rm s}/[(1-\nu)L]$. Equating this to the 
pinning force, one obtains the 'Larkin length' $L_{\rm c} = 
[\mu b[(1+\nu)/(1-\nu)] \sqrt{\xi_{\rm s}/\langle (\sigma^{\rm p})^2\rangle}]^{2/3}$.
The depinning stress can be envisaged as the strength of the pinning field averaged over 
the Larkin length, i.e. 
$\sigma_{\rm c} = \sqrt{\langle (\sigma^{\rm p})^2\rangle
\xi_{\rm s}/L_{\rm c}}$. This estimate yields $\tilde{\sigma}_{\rm c} 
\approx [M\sqrt{C}(1-\nu)/(1+\nu)]^{1/3}$. 

Close to the depinning transition the dislocation is expected to be
rough and satisfy
a self-affinity transformation $h(bx)\sim b^\zeta h(x)$,
where $\zeta$ is the roughness exponent. In practice, we can
obtain $\zeta$ from the power spectrum of $h(x)$, which scales as
$S(k) \sim k^{-(2\zeta+1)}$.
Power counting suggests that the equation for the dislocation dynamics is  
in the same universality class as an equation in which the self stress is 
replaced by $\nabla^2 h$, which corresponds to a line tension approximation. 
Numerical simulations indicate that in the latter case the lineshape is 
superrough with $\zeta \simeq 1.25$ \cite{LES-97}. In addition, transverse 
motion can lead to a fractal lineshape for strong disorder \cite{SEV-91}.

\section{Lattice model}

In order to test the analytical considerations we use a lattice model which allows for 
transverse motion of dislocation segments in the $x$ direction. We discretize the system 
into a grid of square cells with boundaries parallel to the $x$ and $y$ axes and we assign
a ``spin'' $s_i=1$ to the slipped region and $s_i=-1$ to the unslipped region. The 
dislocation is defined as the interface between slipped and unslipped region. By definition, 
this implies that it is discretized in terms of screw and edge segments.

The stress on a given dislocation segment 
is computed by summing up the external stress, the pinning stress and 
the stress due to the other segments which is given by
\beq
\sigma_{xz}^{\rm s}(i)=\sum_j^{(s)}\sigma_s(\vec{r}_i-\vec{r}_j)
+\sum_j^{(e)}\sigma_e(\vec{r}_i-\vec{r}_j)
\eeq
where $\vec{r}_i$ is the point of gravity of the $i$th segment, 
the first sum is over the screw segments 
and the second over the edge ones, $\sigma_s = S_s\mu b^2 x/[4\pi |r|^3]$
and $\sigma_e = S_e \mu b^2 y /[4 \pi (1-\nu)|r|^3$ are the stress fields
due to screw and edge segments in the origin and the signs $S_s$, $S_e$ depend 
on the orientation of the segments. Periodic boundary conditions
are imposed in the $x$ direction by performing the Ewald sum.
In order to avoid possible problems with 
the interaction of near segments \cite{DEV-92}, we 
identify the segment length with the dislocation core radius and neglect
the self-energy of the segments.

The dynamic evolution is performed by chosing at random one segment 
and moving it one step by flipping one spin in the direction of 
the local force. At each time step the stresses are updated. In order 
to test the effects of 
the random update and of the discretization of the dislocation, we 
considered the bowing of a dislocation line pinned between two points
(Fig.~1). This configuration has been investigated 
in a numerical study by Foreman \cite{FOR-67}. In spite of the severe
simplifications we make, the critical stresses and critical configurations
are in good agreement with results of that work. 

We simulate the lattice models using square lattices
of size up to $L=1000$. For the pinning field, we use again a random stress
field created from a Wilkens construction performed by placing two screw
dislocations of opposite sign and line direction normal to the glide plane
at random in a box of size $l$. This corresponds to $C = 1/(4\pi)$, $M = 0.35$. 
In the simulation reported here we further assume $\nu=1/3$ and measure
stresses in units of $\mu$ and distances in units of $b$.
For the dislocation density creating the pinning field, we use the value 
$\rho=0.00125$. 
We obtain $\sigma_c$ by studying the survival probability $P_s$ \cite{KOI-00}
for small lattice sizes $L\simeq 100$. In the limit
$L\to\infty$ one expects $P_s(\sigma)=\theta(\sigma-\sigma_c)$,
where $\theta(x)$ is the Heaviside function. For finite
$L$, we impose that $\sigma_c$ lies in the region for 
which $0.5<P_s<0.7$ and obtain $\sigma_c=0.0040\pm0.0005$
for $\rho=0.00125$ (see Fig.~2). 

For comparison, we note that the Larkin estimate yields 
$\sigma_{\rm c} = 0.0037$. This value is consistent with the simulation
result, although the Larkin length $L_{\rm c}$
is not large as compared to the correlation 
length $\xi_{\rm s}$ as required by the ``weak pinning'' 
assumption.

Simulating the model for larger system sizes, we see
that the dislocation roughens in time but overhangs
and islands are not observed (i.e. the lineshape is
not fractal, see Fig. 3). 
We next compute the power spectrum of $h(x)$ at $\sigma_c$
for a system of size $L=1024$, averaging over ten realizations
of the disorder, and find $\zeta=1$ (see Fig.~4). This result 
differs from  the value $\zeta=1.25$ obtained
using a line tension approximation \cite{LES-97}. 

\section{Discussion and Conclusions}

We have investigated the depinning of a dislocation moving 
in a fixed glide plane and interacting with a random field that has 
long-range correlations. For creating the pinning field, we have used a 
Wilkens construction with an array of immobile 'forest' dislocations 
threading normally through the glide plane. It is emphasized that this 
is {\em not} considered to be a realistic treatment of the interaction 
between moving and forest dislocations -- we have neglected junction 
formation, while the Burgers vector and line direction of the pinning 
dislocations have been chosen in such a manner as to yield maximum 
stresses in the glide plane of the moving dislocation. 
Rather, it was our aim to create a pinning field 
which (i) has correlations with a range of about half the spacing of the
pinning dislocations and (ii) a pinning strength which corresponds approximately 
to the strength of a random forest. (For the present parameters and using 
relations given in \cite{SEV-93}, this amounts to $\sigma_{\rm c} 
\approx 0.0060$.) In this sense the present treatment is supposed to yield a 
more realistic picture of the depinning of a dislocation interacting with a 
dislocation forest than previous models assuming point obstacles \cite{SEV-91}.

As a result, we find that the use of a point obstacle field in conjunction 
with a line tension approximation tends to overestimate the roughness of
the dislocation at the depinning transition. Long range self-interactions give 
rise to a roughness exponent which is smaller than the value obtained from 
a simple line tension approximation. The lineshape of a 
dislocation depinning from a dislocation forest is of significant importance
for theories of work hardening and microstructure evolution. Fractal 
lineshapes of depinning dislocations may help to explain the observation of 
fractal dislocation patterns \cite{SEV-91,HAH-98}, and the fact that a 
moving fractal dislocation must leave behind loops has
 been invoked as a reason for
work hardening \cite{SEV-93}. While the present results do not support
this viewpoint, further studies are required to analyze possible crossovers
to fractal dislocation lineshapes in stronger pinning fields.

\begin{figure}
\centerline{
\psfig{figure=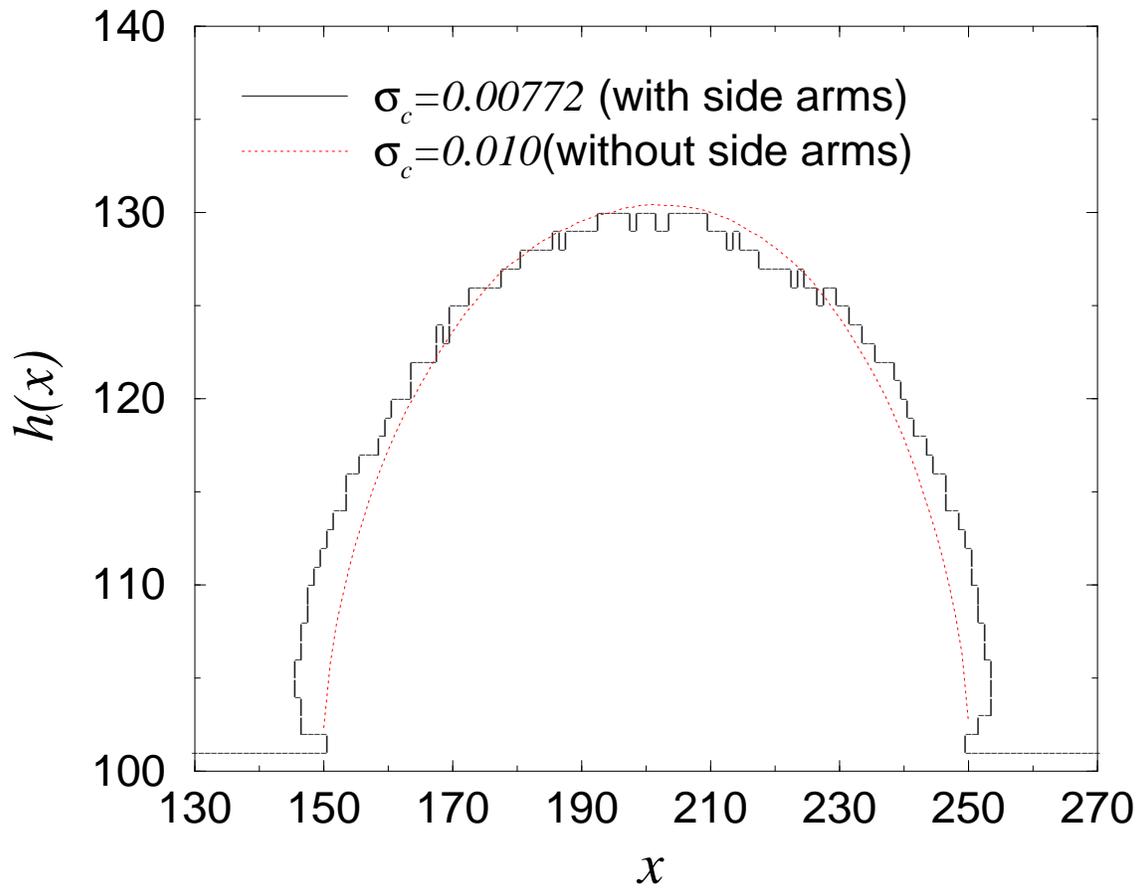,width=15cm,clip=!}
}\caption{Bowing of a dislocation pinned at the two ends, with
at without side arms. For the two cases Ref.~[1] reports
$\sigma_c\simeq 0.0011$ and $\sigma_c\simeq 0.0008$}
\end{figure}

\begin{figure}
\centerline{
\psfig{figure=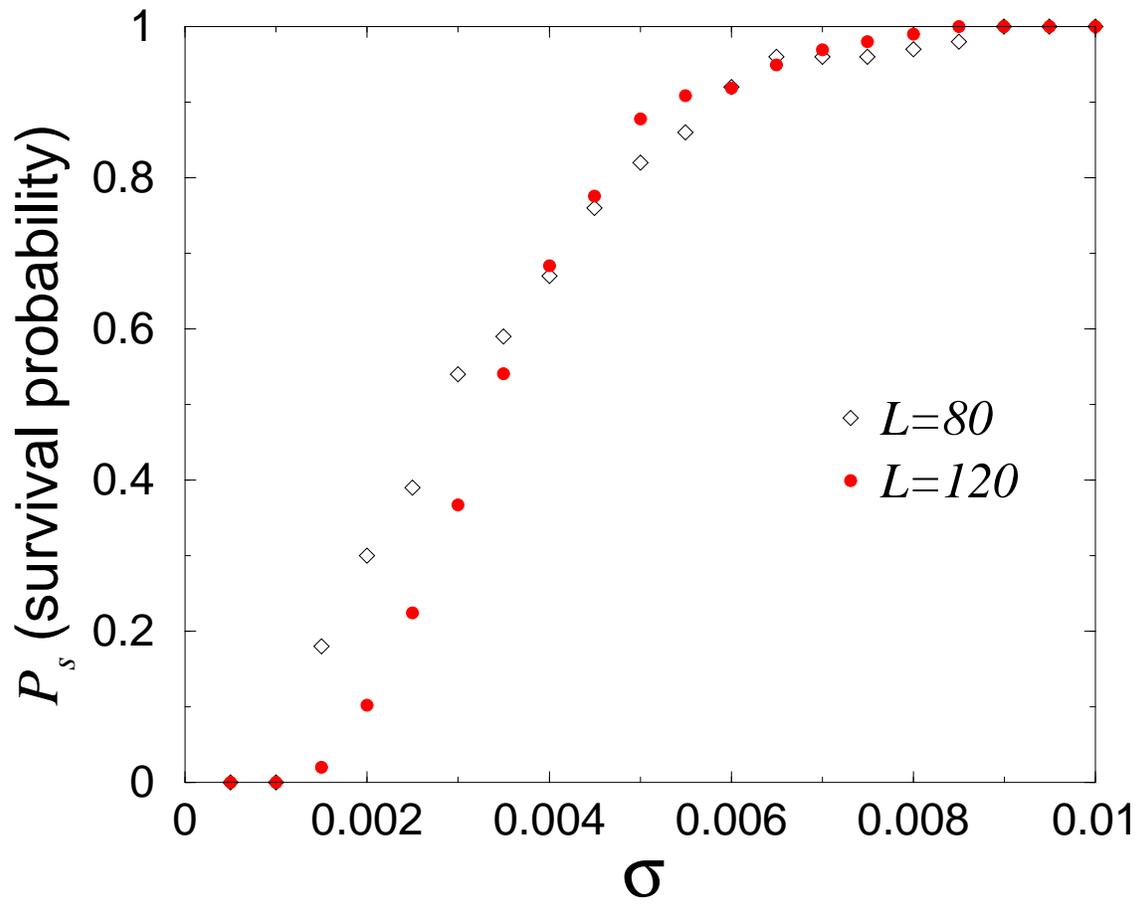,width=15cm,clip=!}
}\caption{ The survival probability for $L=80,120$.}
\end{figure}

\begin{figure}
\centerline{
\psfig{figure=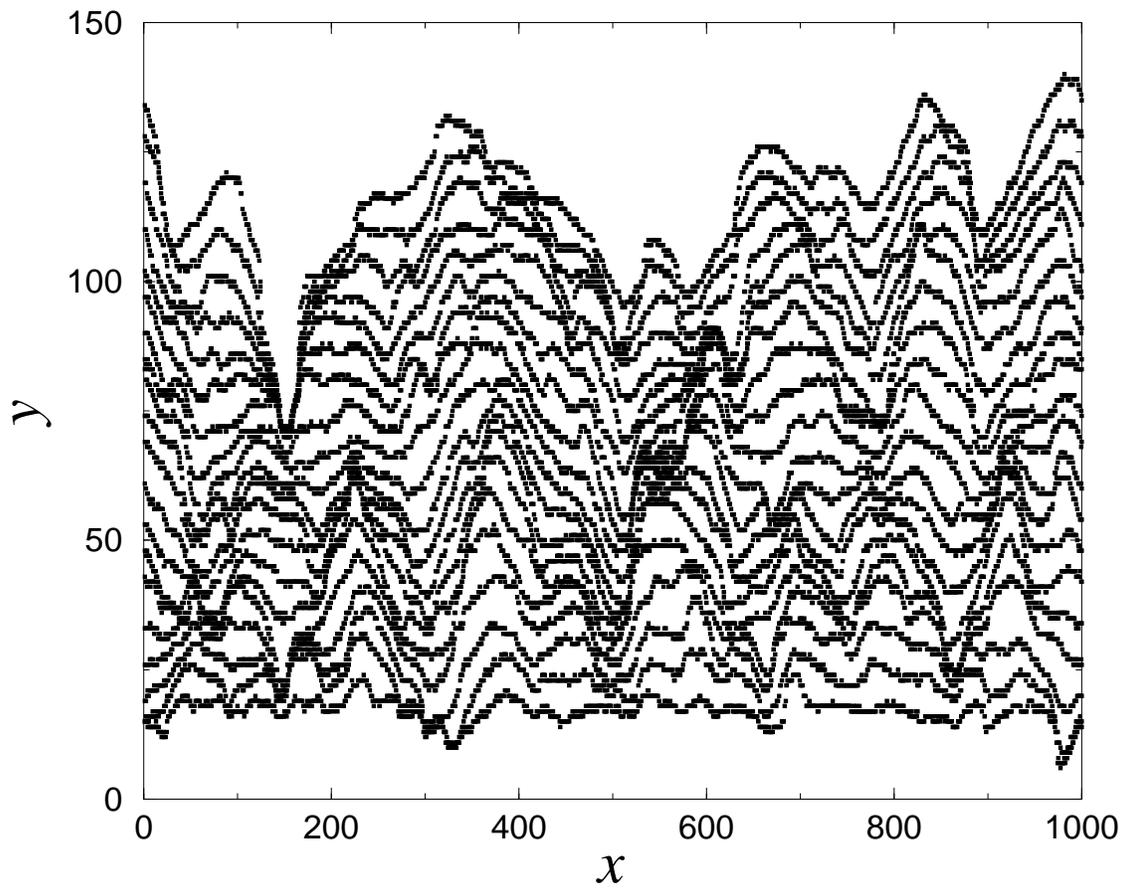,width=15cm,clip=!}
}\caption{ The evolution of the dislocation line for $\sigma=0.005$
and $L=1000$.}
\end{figure}

\begin{figure}
\centerline{
\psfig{figure=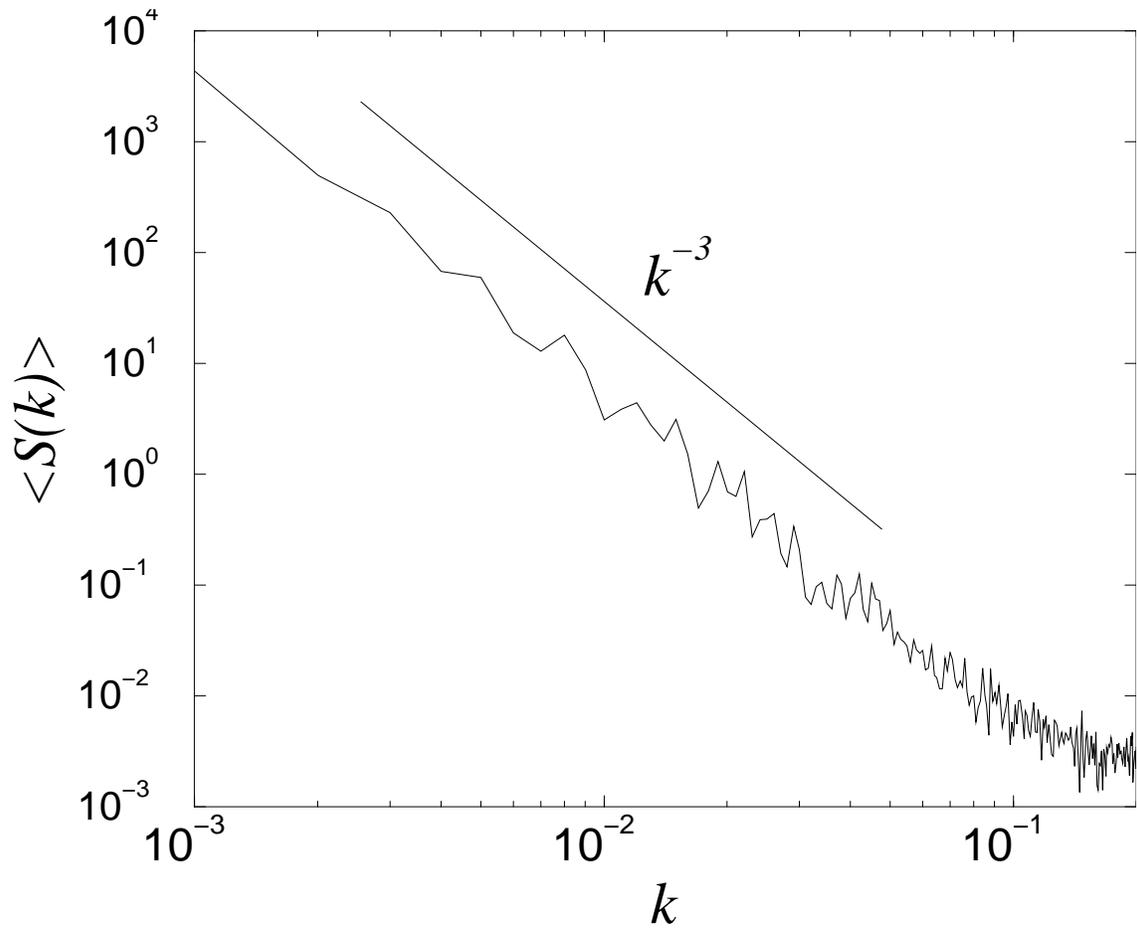,width=15cm,clip=!}
}\caption{ The power spectrum of the interface for $L=1024$ and
$\sigma=0.004$, showing $\zeta \simeq 1$}
\end{figure}

\end{document}